# Voltage-Based State of Charge Correction at Charge-End

Ali Abdollahi, Jianwei Li, Xiaojun Li, Trevor Jones, Asif Habeebullah
*Department of Application Software Design*
*Gotion Inc.*
Fremont, CA USA
Email: t.jones@gotion.com

*Abstract*—A voltage-based method is proposed to correct battery pack state of charge (SOC) estimation at the charge-end. Two main characteristics make the charge-end time span a good opportunity to correct SOC estimation: first, it is easy to detect when the battery is at the last stage of charging because the charging profile is known to the BMS designer and also during the charge-end time span the amount of current is low, and the terminal voltage of the battery cells are high; second, as the battery reaches the charge-end stage, we know that the true SOC is approaching to 100%. This paper presents a method to utilize these important features to correct the SOC estimation error. Using a voltage threshold method, the algorithm detects when the battery is close to the charge-end to activate the charge-end SOC correction strategy. Once activated, the strategy corrects the SOC using the maximum cell voltage to guarantee that SOC is 100% when charging is complete. The amount of correction is a function of maximum cell voltage and the charge current C-rate.

## I. INTRODUCTION

State of charge (SOC) estimation is one of the essential features of a battery management system (BMS). The traditional method of SOC estimation is coulomb counting (or its enhanced versions [1-3]), which is based on calculating accumulated charge injected into or drawn from the battery. There are many other advanced methods for SOC estimation, which can be categorized as model-based approaches or data-driven ones. The model-based methods can be categorized based on the model used, which can be the equivalent circuit model (ECM) or electrochemical model. As for the estimation method, different estimation methods have been used for SOC estimation, including extended Kalman filter (EKF) [4-8], adaptive extended Kalman filter (AEKF) [9-11], unscented Kalman filter (UKF) [12-13], recursive least square [14] and among the data-driven approaches one can refer to long short-term memory (LSTM) [15-17] and deep neural networks [18-19]. The interested reader may refer to review papers in this field [20-21].

Independent of the SOC estimation method utilized, different sources can contribute to SOC estimation error, including initial SOC value, parameters of the SOC estimation algorithm (in case of coulomb-counting method, these include battery capacity and coulombic charge efficiency, and in case of ECM-based estimation methods, these include resistance, capacity, SOC-OCV characterization curve, battery capacity, and coulombic efficiency), measurement error (in case of coulomb-counting method, this is current measurement error, and in case of ECM-based estimation methods, this includes current measurement error and battery terminal voltage measurement error), and battery self-discharge. Therefore, independent of the utilized model or algorithm, it is rational to use any other information which can help correct the SOC estimation and is not already utilized by the SOC estimation method. In this case, since this correction itself requires an algorithm, the final SOC estimation method becomes a hybrid SOC estimation method, composed of the baseline SOC estimation algorithm and SOC correction algorithm.

Two main features make the charge-end time span a good opportunity to correct SOC estimation: first, it is easy to detect the charge end because the BMS designer is responsible for the charging strategy, and hence has a prior knowledge of charge profile and this knowledge can be exploited to robustly predict when the BMS is close to charge-end. In addition, independent of the charging algorithm [22-25], current is at low C-rates as the battery reaches the charge end, which helps make the charge end detection easier. Second, the true SOC at the charge end is known (it is 100%), which helps the BMS developer to gauge the SOC estimation error if there is a big difference between the SOC estimation and 100% once the full charge event is received. This paper presents a method to utilize these important features to correct the SOC estimation.

## II. CHARGE-END STRATEGY FOR SOC CORRECTION

The proposed method for pack SOC correction at the charge-end time-span is voltage-based. It is assumed that once the maximum cell voltage reaches $V_{100\%}$ volts, the pack SOC is 100%. The method is independent of battery chemistry, and with the change of values of the algorithm parameters, it is applicable to any cell chemistry.

The proposed strategy is depicted in Fig. 1, where it is assumed that there exists a baseline SOC estimation method which performs the BMS-reported SOC estimation except during the charge-end period. During the charge-end period, however, the charge-end SOC estimation is activated.

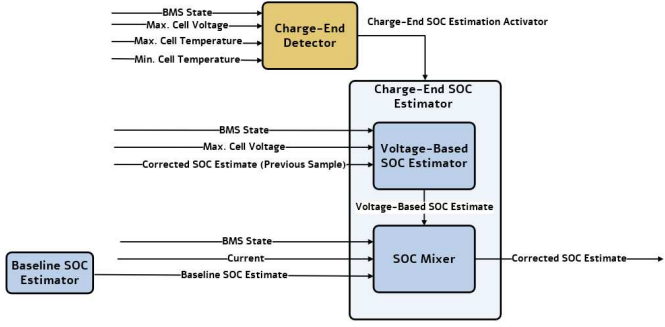

Fig. 1. Charge-end strategy for SOC correction

The strategy is composed of three main blocks, as depicted in Fig. 1 and explained in the following:

- Charge-End Detector: This block monitors the BMS state machine, maximum cell voltage, minimum and maximum cell temperatures, and based on this data, it determines whether the BMS is close to charge-end or not, and if the former is the case, it activates the charge-end SOC correction strategy.
- Voltage-Based SOC Estimator: This block estimates the SOC based on maximum cell voltage and guarantees that the corrected SOC hits 100% when maximum cell voltage reaches $V_{100\%}$ volts.
- SOC Mixer: This block provides the final SOC estimate by mixing the baseline SOC estimate and the voltage-based SOC estimate.

In the following, we describe the detail of these blocks.

*A. Charge-End Detector*

To activate the charge-end strategy for SOC correction, two conditions need to be met, namely:

- BMS is in charge mode
- Maximum cell voltage is greater than a specified threshold ($V_{thr}$) for a specific debounce time ($t_{debounce}$)

The charge-end detector flowchart is depicted in Fig. 2. The output of this subsystem is a Boolean variable indicating either activation or deactivation of the charge-end strategy. At the BMS power-on, the charge-end strategy is deactivated, and it remains deactivated in any non-charging (not AC charging nor DC charging) BMS state.

During DC charging mode, if the maximum cell voltage is greater than a specific threshold for a specific amount of time (debounce time), then the charge-end strategy is activated. The threshold voltage is a function of minimum and maximum cell temperatures. This function can be a fixed voltage threshold, or a linear or nonlinear regression of minimum and maximum cell temperatures.

The DC threshold mapping is designed using the DC charge profiles that have ended up with full charge (i.e., final SOC of 100%). For each profile, the SOC is calculated using backward coulomb-counting, where the initial time is the moment of reaching charge-end and SOC at the initial time is assumed to be 100%. The SOC of all other time samples before charge-end are calculated by progressing time reversely starting from the initial time (i.e., the moment of reaching charge-end). Then, a suitable SOC threshold (e.g., 80%) is selected for activating the charge-end strategy; this SOC threshold is called $soc_{threshold}$.

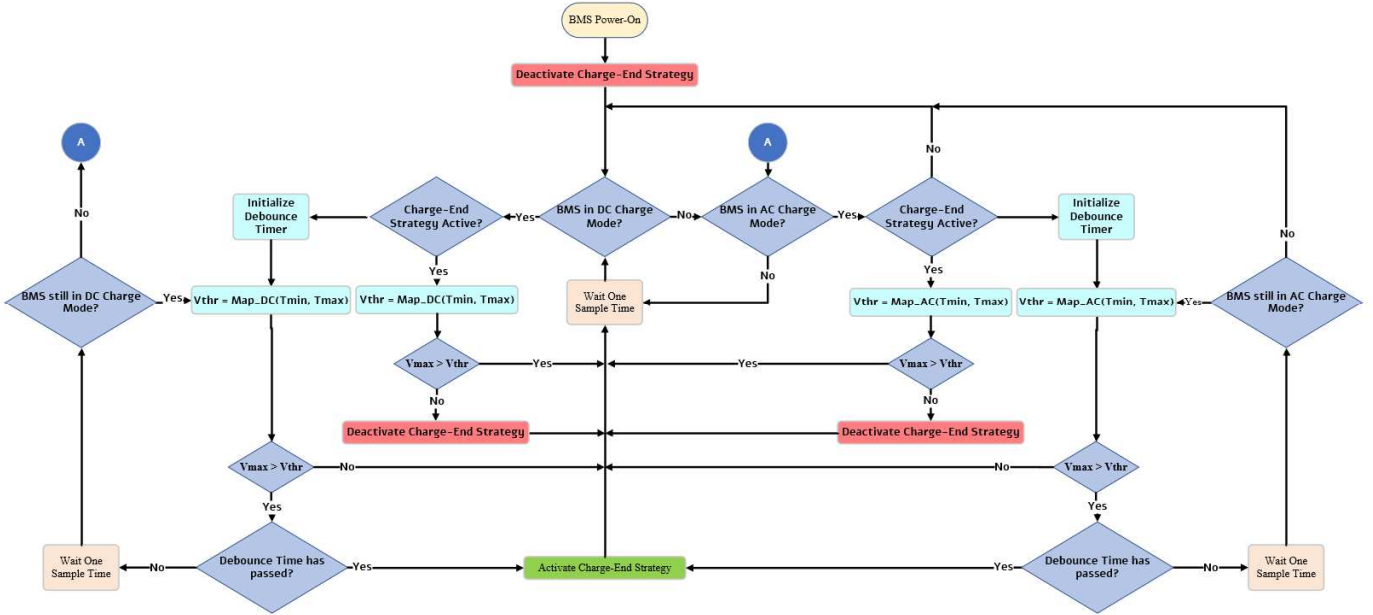

Fig. 2. Flowchart of charge-end detector

For each DC profile, the maximum cell voltage, minimum cell temperature, and maximum cell temperature are obtained at the time that backward-calculated SOC hits $soc_{threshold}$.

Alternatively, one can use a $TTC_{threshold}$ (a time-to-charge (TTC) threshold before reaching full charge) instead of $soc_{threshold}$. For example, by selecting $TTC_{threshold} = 30\ min$, for each profile, the maximum cell voltage, minimum cell temperature, and maximum cell temperature are obtained at 30 minutes before the charge process completes.

Once such data is gathered for all training sets of DC charge profiles, the DC threshold function is devised using an appropriate training algorithm (e.g., linear regression). More advanced, data-driven approaches can be used for determining the threshold mapping.

A similar logic is applicable to activating charge-end strategy in AC charging mode, and a similar approach is used to devise a threshold mapping for maximum cell voltage; however, the threshold mappings for AC mode would be different from that of the DC mode because the characteristics of the AC and DC charging strategies are different.

An equivalent logic diagram for charge-end strategy is shown in Fig. 3. Here it is assumed that IsDC_ChargeMode and IsAC_ChargeMode are Boolean signals, where IsDC_ChargeMode is true when BMS is in DC state mode, and IsAC_ChargeMode is true when BMS is in AC state mode. Also, the "Debouncer" block is a block that its output becomes true if its input remains true for a specific debounce time ($t_{debounce}$), and its output drop to false when its input change to false.

### B. Voltage-Based SOC Estimator

The voltage-based SOC estimation is based on the assumption that SOC equals 100% when maximum cell voltage reaches $V_{100\%}$. At time $k$, since the maximum voltage is known and corrected SOC and maximum cell voltage are known at one sample earlier, i.e., at time $k-1$, therefore, we can find the voltage-based SOC estimate using linear interpolation as shown in Fig. 4.

For this purpose, we can calculate the rate of change of SOC over the change of maximum voltage between points B and A and equate it with that of points C and A, as follows:

$$\frac{soc_{VB}(k)-soc_{corr}(k-1)}{v_{max}(k)-v_{max}(k-1)} = \frac{100-soc_{corr}(k-1)}{V_{100\%}-v_{max}(k-1)_{corr}} \quad (1)$$

The above equation can be solved for $soc_{VB}(k)$ as follows:

$$soc_{VB}(k) = soc_{corr}(k-1)$$
$$+\frac{v_{max}(k)-v_{max}(k-1)}{V_{100\%}-v_{max}(k-1)}(100-soc_{corr}(k-1)) \quad (2)$$

Note that $soc_{corr}(k-1)$ at the start of strategy is initialized with $soc_{baseline}$ (which is the SOC value estimated by the baseline estimator). The above equation can be improved by adding an acceleration parameter of $\gamma \geq 1$, as follows, provided that a limiter be applied to $soc_{VB}(k)$ so that it does not exceed 100%.

$$soc_{VB}(k) = soc_{corr}(k-1)$$
$$+\gamma\frac{v_{max}(k)-v_{max}(k-1)}{V_{100\%}-v_{max}(k-1)}(100-soc_{corr}(k-1)) \quad (3)$$

It is reasonable to have such a limiter even if $\gamma = 1$, because measurements greater than $V_{100\%}$ are possible, which cause final calculated SOC be greater than 100%. Therefore, for any value of $\gamma \geq 1$, the following equation guarantees that $0 \leq soc_{VB}(k) \leq 100$.

$$soc_{VB}(k) = \min\left\{100, \max\left(0, soc_{corr}(k-1) + \gamma\frac{v_{max}(k)-v_{max}(k-1)}{V_{100\%}-v_{max}(k-1)}(100-soc_{corr}(k-1))\right)\right\} \quad (4)$$

### C. SOC Mixer

The last stage of the charge-end strategy for SOC correction is mixing the voltage-based SOC estimate and the baseline SOC estimate.

Fig. 5 shows the flowchart of the mixing algorithm. The output of mixing algorithm ($soc_{corr}$) is a linear combination of voltage-based SOC estimate ($soc_{VB}$) and baseline SOC estimate ($soc_{baseline}$) as follows:

$$soc_{corr} = \alpha\ soc_{VB} + (1-\alpha)\ soc_{baseline} \quad (5)$$

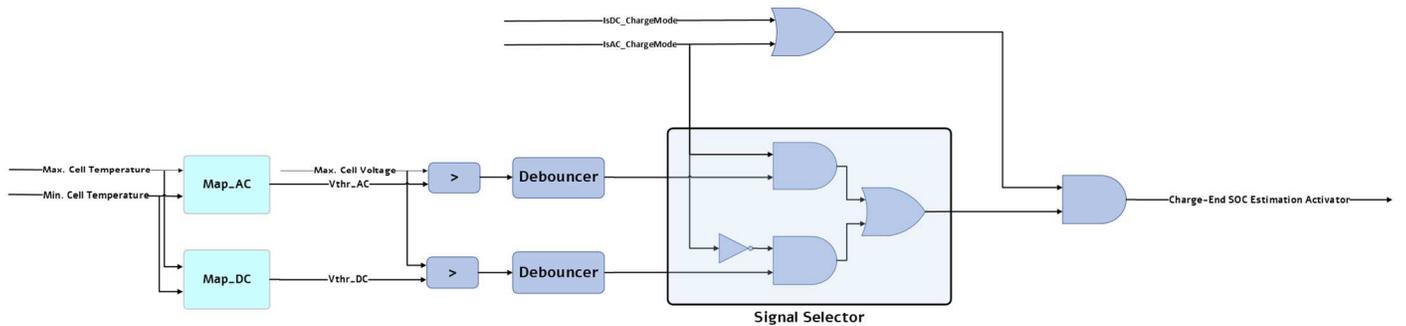

Fig. 3. Logic diagram of the charge-end detector

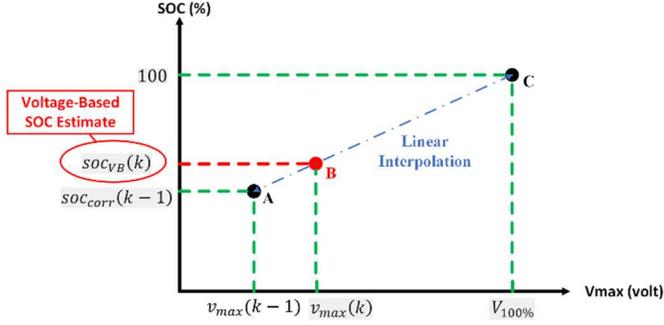

Fig. 4. Voltage-based SOC estimation

However, this is the case only when the voltage based SOC estimate is higher than the baseline SOC estimate; otherwise, no correction happens; that is

$$soc_{corr} = soc_{baseline} \qquad (6)$$

This criterion is adopted because, at the charge end, it is against user experience to reduce the SOC estimate. Thus, we propose only increasing SOC, if any, as the correction at the charge end.

The mixing coefficient $\alpha$ is a tuning factor in the range of [0, 1] that needs to be determined based on charge profiles; therefore, in general, it is different for AC and DC charge modes.

Fig. 6 shows two examples of this mixing coefficient. An important characteristic of this coefficient is that the lower the C-rate of the current, the higher the value of the coefficient, resulting in more voltage-based SOC correction. This is because as the charging approaches to its end (full charge), the magnitude of charge current generally decreases.

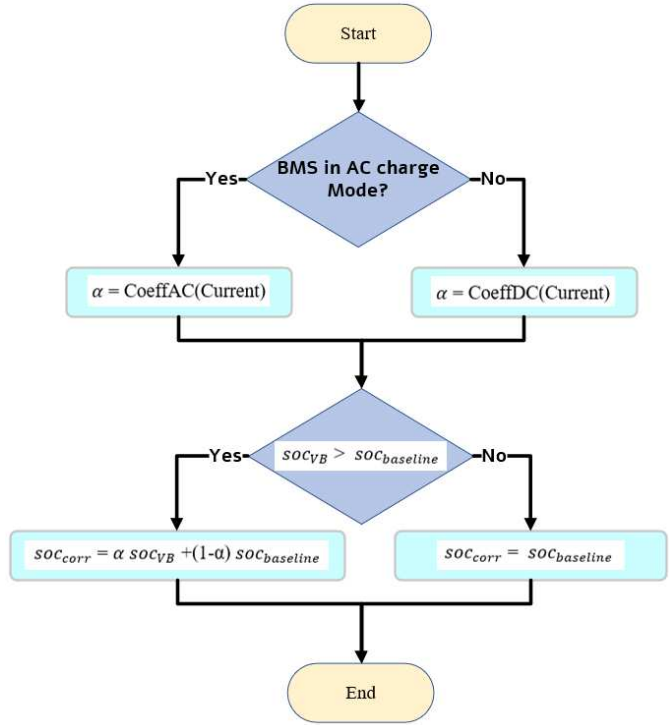

Fig. 5. Mixing voltage-based SOC estimate and baseline SOC estimate

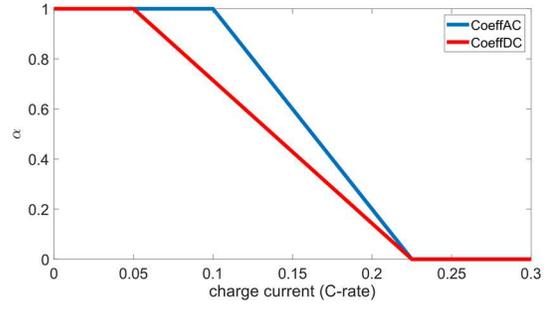

Fig. 6. Examples of mixing coefficient of $\alpha$ as a function of charge current (C-rate) for AC and DC profiles

## III. SIMULATION RESULTS

In this section, we discuss how the proposed method works using simulation. For this purpose, the field data obtained from several vehicles are used as inputs of the simulation. The baseline algorithm is chosen as a coulomb-counting method that jumps to SOC of 100% when the maximum cell voltage hits $V_{100\%}$. In order to observe the effect of the charge-end SOC correction strategy, in all simulations, the baseline algorithm is initialized with a wrong SOC value. In all simulations, the time that the battery reaches full charge is considered as reference and is shown as zero, and any time before that is shown as a negative number in seconds.

Fig. 7 shows a scenario where the SOC is initialized with more than 15% error and a low threshold voltage is used in the charge-end detector subsystem. As it is seen, about an hour before charge-end, the corrected SOC is increasing more than the coulomb-counting calculation, and as the time reaches the charge-end moment, this SOC rise becomes more forceful. Note that as the charging process approaches its end, the corrected SOC smoothly approaches 100%.

In Fig. 8, the simulation is repeated with a higher threshold value for maximum cell voltage. The figure shows that the charge-end SOC correction strategy starts about a quarter an hour later. However, the corrected SOC still smoothly increases and approaches 100% as the charge process reaches its end.

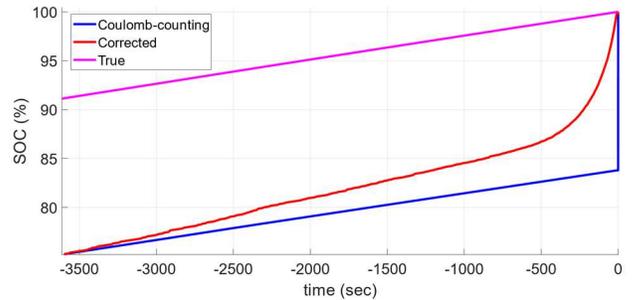

Fig. 7. Comparison of the corrected strategy with a low threshold value for maximum cell voltage

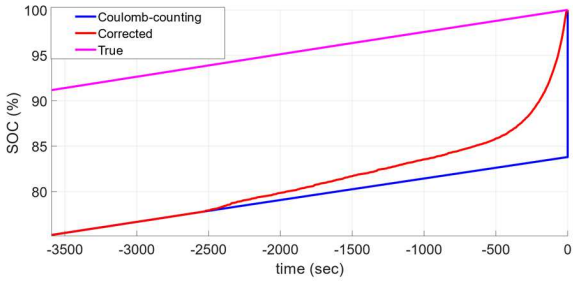
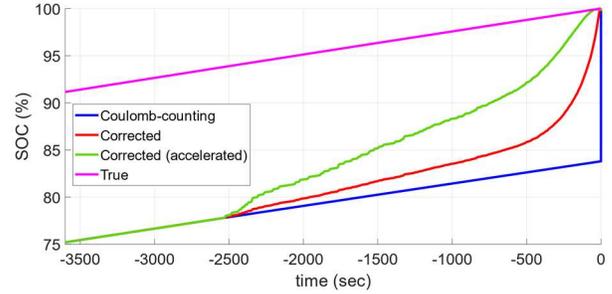

Fig. 8. Comparison of the corrected strategy with a high threshold value for maximum cell voltage

Fig. 9. Comparison of the corrected strategy with a high threshold value for maximum cell voltage, with two different values (one equal to l the other greater than 1) for acceleration term

A problem with the previous simulations is that most of the correction happens in the last 10 minutes, and this causes SOC to increase too much in this short interval which is not good from a user-experience point of view. To resolve this issue, we can use a greater than 1 value for the acceleration parameter of $\gamma$ (in the previous simulations, this value was set as 1). The result of this simulation is shown in Fig. 9. As it can be seen, using a greater than 1 value for the acceleration parameter of $\gamma$ helps distribute the SOC correction over a much wider time frame before the battery reaches the full-charge state.

Fig. 10 shows the simulation result for nine different sets of field data used as inputs of the proposed strategy. Here for each profile, four different simulations are run by injecting 5%, 15%, 25%, and 35% initial SOC error into the baseline coulomb-counting SOC calculation. As it is seen the SOC error of the corrected SOC smoothly approaches zero as the charge profile approaches its end state.

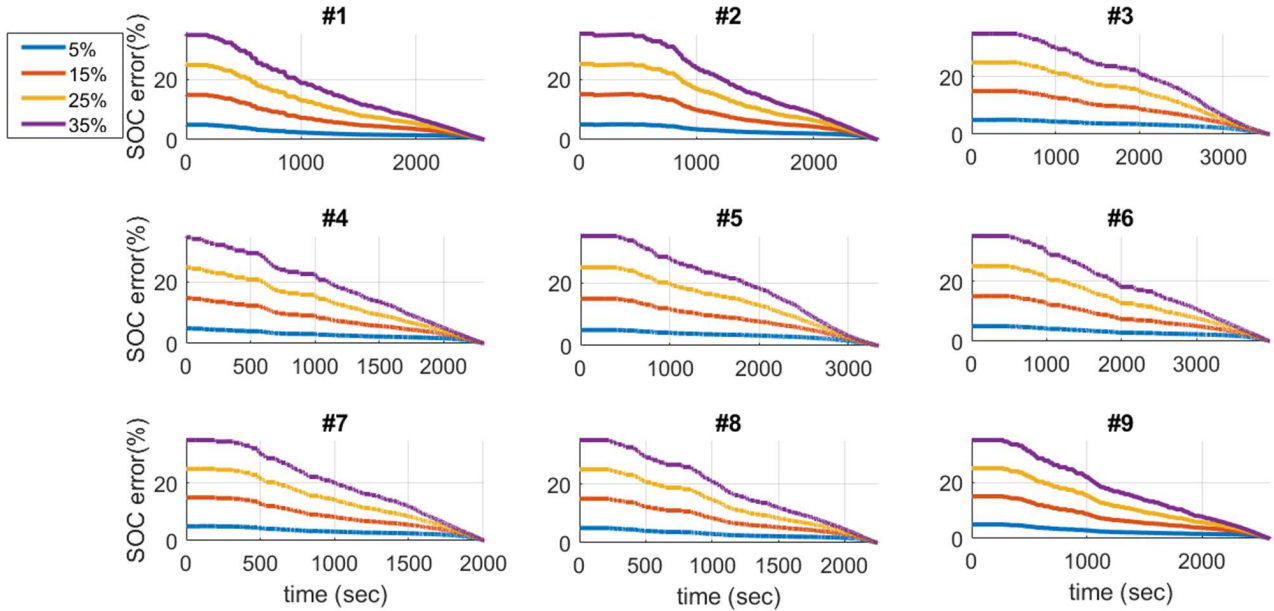

Fig. 10. SOC error of charge-end strategy when different initial errors of (5%, 15%, 25, 35%) are introduced for nine different sets of data

## IV. CONCLUSIONS AND FUTURE WORK

In this paper, we presented a voltage-based SOC correction algorithm to be used at the charge-end time span to correct the SOC estimation performed by the baseline SOC estimator. The method is composed of three subsystems: charge-end detector, voltage-based SOC estimator, and SOC mixer. The charge-end detector subsystem uses the BMS state mode and maps minimum and maximum cell temperatures to a maximum cell voltage threshold. It detects charge-end once the maximum cell voltage is greater than the specified threshold for more than a specified debounce time. The voltage-based SOC estimator estimates the SOC based on maximum cell voltage and guarantees that the corrected SOC reaches 100% when maximum cell voltage hits $V_{100\%}$ volts. The SOC mixer subsystem uses a linear combination of voltage-based SOC estimate and the baseline SOC estimate. The linear coefficient parameter is a function of charge current, and as the charge current reduces (while the charge process approaches its end), this coefficient becomes larger which results in weighting the voltage-based SOC estimate more. We demonstrated the effectiveness of the proposed method by initializing the coulomb-counting method (which was chosen as baseline estimator in this paper) with a wrong initial value. The proposed strategy was capable of detecting the charge-end and correcting the estimated SOC so that it smoothly approaches 100% as the charge profile reaches its end.